# Superconductivity at 3.7 K in Ternary Silicide Li$_2$IrSi$_3$

Daigorou Hirai[1], Rui Kawakami[1], Oxana V. Magdysyuk[2], Robert E. Dinnebier[2], Alexander Yaresko[2] and Hidenori Takagi[1,2]

[1]*Department of Physics, University of Tokyo, 7-3-1 Hongo, Bunkyo, Tokyo 113-0033, Japan*

[2]*Max Planck Institute for Solid State Research, Heisenbergstrasse 1, D-70569 Stuttgart, Germany*

We report the discovery of superconductivity at $T_c$ = 3.7 K in the new ternary lithium silicide Li$_2$IrSi$_3$. The crystal structure of Li$_2$IrSi$_3$ consists of IrSi$_6$ antiprisms connected by Si triangles, giving rise to a three dimensional framework of covalent Si-Si and Si-Ir bonds. Electronic specific-heat in superconducting phase suggests that Li$_2$IrSi$_3$ is a BCS weak-coupling superconductor.

The discovery of superconductivity in MgB$_2$ with a transition temperature ($T_c$) as high as 39 K [1] revived interests in phonon-mediated superconductivity as one of roads to high $T_c$. MgB$_2$ belongs to a class of 'covalent metal' [2], where charge carriers are in strongly directional covalent bonds. The covalent character may play a vital role in realizing phonon-mediated superconductors with a high $T_c$. In the Macmillan formula [3], $T_c$ is related to an average phonon frequency $<\omega>$, the electron phonon coupling parameter $\lambda_{ep}$ and the Coulomb pseudo potential $\mu^*$: $T_c = (h<\omega>/1.2k_B) \exp\{[-1.04(1+\lambda_{ep})/[\lambda_{ep} - \mu^*(1+0.62\lambda_{ep})]]\}$ The strong covalent bonding giving rise to large phonon frequencies, which merits in enhancing $T_c$ but can suppress $\lambda_{ep} = [N(E_F) <I^2>] / [M <\omega^2>]$ [3]. If the conduction electrons are in the bonds with strong covalent character, however, the suppression of $\lambda_{ep}$ may be minimized through the enhanced $<I^2>$, the electronic matrix element of the change in crystal potential.

Such working hypothesis may be justified by the discovery of superconductivity in boron-doped diamond [4], silicon [5] and silicon carbide [6,7]. In those covalent metals, however, $T_c$ seems to be somehow limited by the difficulty in introducing enough numbers of carriers, namely the solubility of dopants. Superconductivity in doped silicon clathrates [8–10] may give us a hint to achieve heavy doping in covalent metals. Silicon clathrates have a cage-like three dimensional structure made of face-sharing Si clusters as building blocks. Each silicon atom has a local $sp^3$ tetrahedral coordination as seen in the diamond lattice of pure silicon. The characteristic cage-like structure allows large amount of doping by intercalating dopant at the center of the cages with maintaining covalent character of silicon.

Based on the outlook above, we have been searching for new superconductors in doped covalent metals with cage-like framework. Our strategy has been to introduce transition-metals as the third component, which enlarges the phase space of materials for exploration. Transition-metals often form covalent bonds with silicon, since the electronegativity of transition-metals is comparable to that of silicon. In addition, tunability of chemical potential by changing transition-metals may provide us with additional channel in designing the ideal electronic structure.

In this paper, we report new ternary transition metal silicide $Li_2IrSi_3$, which shows superconductivity below 3.7 K. The crystal structure of $Li_2IrSi_3$ consists of Si triangles connected by Ir atoms, giving rise to a three dimensional network of covalent bonds. We argue that the presence of high frequency phonons result in moderate $T_c$ of 3.7 K despite of very clear weak-coupling character.

Polycrystalline samples of $Li_2IrSi_3$ were synthesized by conventional solid-state reaction. Elemental silicon and iridium were mixed and pelletized with small pieces of lithium metal. The pellet was placed in an arc-welded titanium tube and sintered at 850 °C for 48 h in a carbon coated quartz tube filled with a partial pressure of Ar gas. X-ray powder diffraction data of $Li_2IrSi_3$ at 293 K were collected with the wavelength of 0.3998(2) Å at the high resolution powder diffractometer at beamline ID31 of the European Synchrotron Radiation Facility (ESRF). The sample of $Li_2IrSi_3$ was placed in a 0.3 mm lithium borate glass capillary, which was rotated around $\theta$. The diffraction data were collected continuously from 1° to 60° for $2\theta$ and rebinned to steps of 0.003° $2\theta$. Magnetic, transport and thermal measurements were performed by using a magnetic properties measurement system (MPMS; Quantum Design) and a physical property measurement system (PPMS; Quantum Design) equipped with $^3$He option. The electronic structure of $Li_2IrSi_3$ was calculated using the fully relativistic Linear Muffin-Tin Orbital method as implemented in the PY-LMTO computer code. Some details of the implementation can be found in Ref. [11].

In the course of experiments of Li-Ir-Si systems, a trace of superconductivity around 3.7 K was observed in mixed phase samples containing $IrSi_3$ and unknown phases. The superconducting phase was identified as $Li_2IrSi_3$ by changing the starting composition of Li-Ir-Si and by comparing XRD patterns and the magnitude of the diamagnetic signal of superconductivity. The single phase sample of $Li_2IrSi_3$ has metallic silver color and was stable in air. The XRD pattern for $Li_2IrSi_3$ can be indexed with the hexagonal space group *P*6$_3$/*mmc* [Fig. 1(a)]. None of the known crystal structures in the databases could have reproduced the XRD pattern of the purified phase. Therefore, the crystal structure of $Li_2IrSi_3$ was solved independently by simulated annealing and by charge flipping [12–14] using the tools of the program TOPAS 4.2 [15]. Both structure determination techniques resulted in identical structural models. Subsequently, Rietveld refinement was performed with freely refining atomic positions, anisotropic ADPs, and the occupancy of the Li atom. Off-stoichiometry of Li was not detected by XRD refinement within the error of 10%, and the occupancy of Li was thus fixed to unity for the final refinement. The refinement converged quickly. Structural parameters for $Li_2IrSi_3$ are summarized in Table I. Stoichiometric composition of Li:Ir:Si = 2:1:3 was double checked by chemical analysis by inductively coupled plasma optical emission spectroscopy.

The crystal structure of $Li_2IrSi_3$ is composed of $IrSi_6$ antiprisms along *c*-axis, which are connected by Si triangles, as illustrated in Fig. 1(b). In the plane perpendicular to the *c*-axis, Si triangles form Kagome lattice with short (2.435(3) Å) and long (2.583(3) Å) Si-Si distances [Fig. 1(c)]. Comparing the Si-Si bond length with those of elemental Si (2.35 Å) [16] and the clathrate $Ba_8Si_{46}$ (2.27 ~ 2.48 Å) [8], shorter bonds in $Li_2IrSi_3$ certainly form covalent bonding, but not longer ones. Looking down the crystal structure along *c*-axis, Li ions are located at the center of hexagonal holes of the Kagome lattice 2.809(7) Å away from nearest Si ions, whereas Ir ions are located at the center of the larger Si triangle with a relatively short Ir-Si distance of 2.4627(9) Å. The strong Ir-Si bonds forming $IrSi_6$ antiprisms and the covalent Si-Si bond form a three dimensional rigid network as a whole.

$Li_2IrSi_3$ was found to show superconductivity below 3.7 K, as evidenced by a large Meissner signal and a zero resistance [Fig. 2(a) and (b)]. The large Meissner fraction (field cool magnetization), about 80% of the perfect diamagnetism, is the hallmark of bulk superconductivity. The field dependence of isothermal magnetization exhibits typical type-II superconductor behavior, as shown in the inset of Fig. 2(a). The upper critical field $\mu_0 H_{c2}(T)$ has been determined by field sweep at various temperatures [inset of Fig. 2(b)]. $T_c$ defined as a mid-point of the resistive transition is systematically suppressed under the magnetic field and saturates toward $\mu_0 H_{c2}(0) \sim 0.21$ T at the zero temperature limit. From $H_{c2}(0)$, the Ginzburg-Landau coherence length is estimated in the orbital limit to be $\xi_{GL}(0) \sim 400$ Å. The very long coherence length suggests this superconductor is relatively clean.

Further support for the bulk nature of superconductivity in Li$_2$IrSi$_3$ was obtained from the large specific heat jump at $T_c$, as shown in Fig. 2(c). Considering the entropy conservation at $T_c$, bulk $T_c = 3.70$ K is determined. Normal state specific-heat was evaluated by suppressing superconducting phase with a magnetic field of $\mu_0 H = 1$ T [inset of Fig. 2(c)]. The fitting of normal state specific-heat data below 7 K with $C_N(T) = \gamma T + \beta T^3$ yields an estimation of $\gamma = 5.73$ mJ / (mol·K$^2$) and $\beta = 0.101$ mJ / (mol·K$^4$). The obtained specific-heat coefficient $\gamma$ is moderately low, as expected for a 5$d$ intermetallic compound. The Debye temperature $\Theta_D = (12\pi^4 NR/5\beta)^{1/3}$ is calculated to be 486 K, where $N$ and $R$ are the number of atoms per formula unit and gas constant, respectively. As expected from the crystal structure composed by Si-Si and Si-Ir covalent bonds, this value is significantly higher than other elemental or intermetallic superconductors, such as Pb ($\Theta_D = 105$ K, $T_c = 7.2$ K) [17], Nb$_3$Sn ($\Theta_D = 234$ K, $T_c = 18$ K) [18], and Ba$_{0.55}$K$_{0.45}$Fe$_2$As$_2$ ($\Theta_D = 230$ K, $T_c = 30$ K) [19]. Surprisingly, the Debye temperature of Li$_2$IrSi$_3$ is even higher than that of superconducting clathrate Ba$_8$Si$_{46}$ ($\Theta_D = 370$ K, $T_c = 8$ K) [10], where rigid $sp^3$ Si-Si bonds form three-dimensional network.

The electronic contribution of specific-heat $C_e/T(T)$ indicates a weak-coupling superconductivity. The normalized specific-heat jump at $T_c$, $\Delta C/\gamma T_c \sim 1.40$ is close to the BCS weak coupling limit value 1.42. Exponential decay of $C_e/T(T)$ below $T_c$ clearly indicates an $s$-wave superconducting gap in Li$_2$IrSi$_3$. $C_e/T(T)$ can be well fitted by an exponential function $C_e/T(T) = A \exp(-\Delta_0/k_B T)$, the so-called $\alpha$ model [20], where $k_B$ and $\Delta_0$ are the Boltzmann constant and superconducting gap at 0 K, respectively. The obtained coupling strength $\alpha = \Delta_0/k_B T_c = 1.63$ is again very close to the value $\alpha = 1.76$ expected for the BCS weak coupling limit. Together with the obtained Debye temperature, $T_c$ and an assumption of $\mu^* = 0.13$ for a typical metal yield the estimation of electron-phonon coupling constant $\lambda_{ep}$ from the McMillan formula. [3] The obtained value $\lambda_{ep} = 0.52$ is comparable to that of weak coupling superconductor, such as aluminum $\lambda_{ep} = 0.38$ [21], and consistent with the weak coupling behavior observed in the specific-heat data.

The calculated band-structure shows rather strong hybridization between Si($3s$ & $3p$) and Ir ($5d$) orbitals around Fermi energy ($E_F$). Orbitally resolved density of states (DOS) obtained from a scalar relativistic calculation for Li$_2$IrSi$_3$ is shown in Fig. 3. Around $E_F$, $3s$ and $3p$ states of Si and $5d$ state of Ir have dominant contribution. This strong hybridization between Si ($p_x$, $p_y$) and Ir ($d_{xz}$, $d_{yz}$) leads to covalent Ir-Si chemical bonds forming the rigid structure of Li$_2$IrSi$_3$. Although spin-orbit coupling of the $5d$ states of heavy Ir is quite strong, it is much smaller than the Ir $5d$ band width and has only weak effect on the band structure. It lifts the degeneracy of some bands crossing $E_F$ but does not affect the chemical bonding and DOS at $E_F$. The band DOS at $E_F$, $N(E_F)$ obtained above corresponds to an electronic specific coefficient $\gamma_{band} = \pi^2 k_B^2 N(E_F)/3 = 3.5$ mJ / (mol·K$^2$). The comparison with the experimentally obtained $\gamma = (1+\lambda_{ep})\gamma_{band} = 5.73$ mJ / (mol·K$^2$) yields an estimation of electron-phonon coupling constant $\lambda_{ep} = 0.63$, which agrees reasonably well with the estimation based on MacMillan formula.

All the results obtained so far point that Li$_2$IrSi$_3$ is a weak-coupling superconductor with relatively high phonon frequency associated with the presence of the covalent bonds. $\lambda_{ep} = [N(E_F) <I^2>] / [M <\omega^2>]$ remains small because of the moderate density of states $N(E_F)$ (and likely $<I^2>$), and the large $<\omega>$. The small $\lambda_{ep}$ is compensated to a certain extent by the high phonon frequency $<\omega>$, which very likely gives rise to not very low $T_c$ close to 4 K.

One of the obvious strategies to enhance $T_c$ further may be to increase density of states $N(E_F)$. If the number of $d$-electrons were decreased by substituting Os for Ir, $E_F$ would shift to lower energy, where the peak of DOS locates. Our naive expectation from the calculation is that Li$_2$OsSi$_3$ would have higher DOS at $E_F$, and hence $T_c$. So far, we could have synthesized two new iso-structural compounds, Li$_2$RhSi$_3$ and Li$_2$PtSi$_3$. They should have the same and one less $d$-electron and therefore lower or comparable $N(E_F)$ respectively, as compared with Li$_2$IrSi$_3$. A superconducting transition was observed at 2.2 K for Li$_2$RhSi$_3$ while no superconducting signal was observed for Li$_2$PtSi$_3$ above 1.8 K. This is consistent with the naive expectation from the comparison of DOS.

In conclusion, we have discovered a new transition metal silicide,

Li$_2$IrSi$_3$, which shows superconductivity at 3.7 K. It is a weak-coupling superconductor with a high phonon frequency associated with the strongly covalent Ir-Si and Si-Si bonds. If electron-phonon coupling constant $\lambda_{ep}$ was optimized by increasing $N(E_F)$ for example, we might able to enhance $T_c$ further with the help of the high phonon frequency.

### Acknowledgement

Authors acknowledge Dr. Ch. Drathen (ESRF, Grenoble) for support with synchrotron measurements (proposal ch3878). This work was supported by a Grant-in-Aid for Scientific Research (No. 24224010) from MEXT, Japan.

### Note

We recently became aware of a report by Pyon et al., on the discovery of superconductivity in Li$_2$IrSi$_3$. [22]

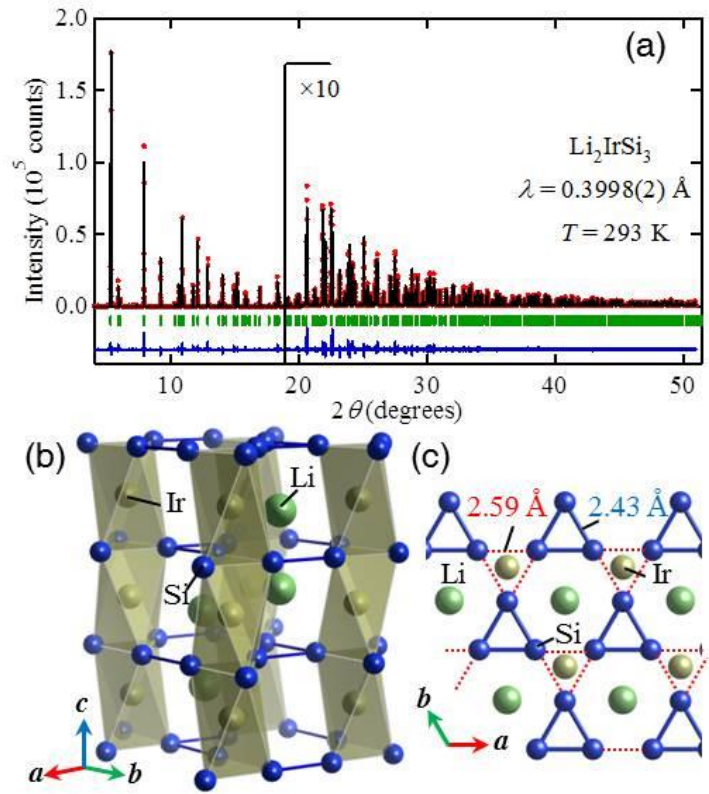

Fig. 1 (a) Scattered x-ray intensities of $Li_2IrSi_3$ as a function of diffraction angle $2\theta$. The observed pattern (circles) measured in Debye-Scherrer geometry, the best Rietveld fit profiles (upper line), peak positions (tick marks) and the difference curve between the observed and the calculated profiles (lower line) are shown, respectively. The high angle part starting at 20.2° $2\theta$ is enlarged for clarity. (b) Crystal structure of $Li_2IrSi_3$. (c) Kagome network of Si on *ab*-plane. Short and long bonds are shown in a different way.

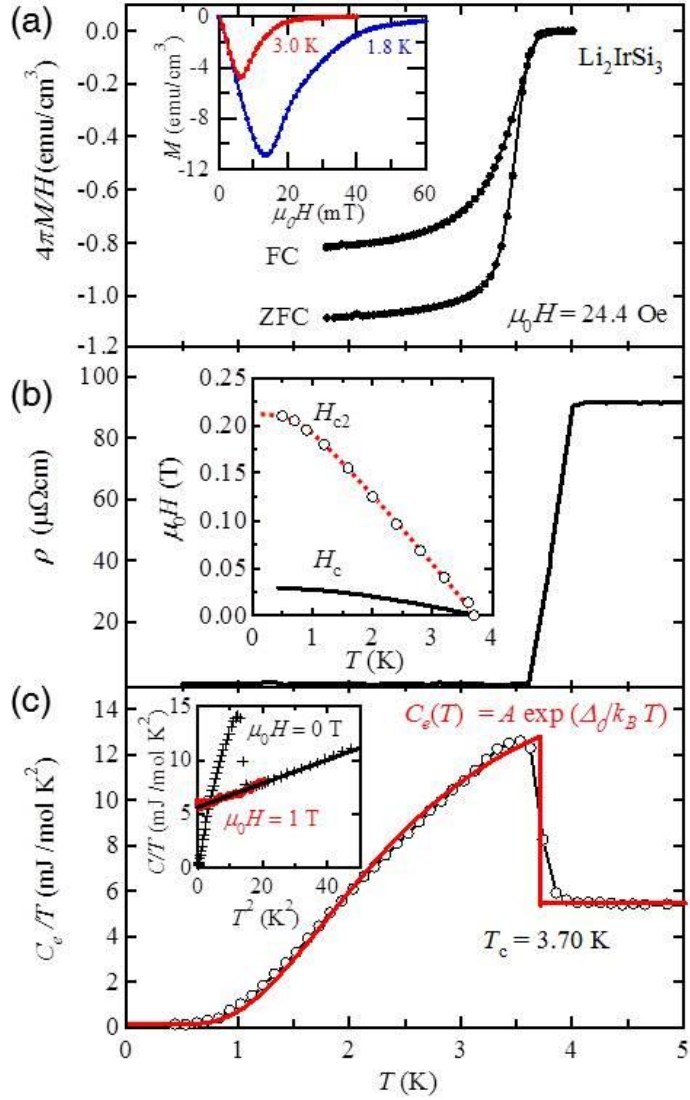

Fig. 2 Temperature dependence of (a) magnetization, (b) resistivity and (c) electronic specific-heat showing superconducting transition of $Li_2IrSi_3$. Insets: (a) Field dependence of magnetization. (b) Temperature dependence of upper critical field $H_{c2}(T)$ (circle) and thermodynamic critical field $H_c(T)$ (solid line) obtained from magnetoresistive and specific-heat data, respectively. The dotted line is a guide to the eyes. (c) $C/T$ vs $T^2$ plot below 7 K under applied field of $\mu_0H = 0$ (cross) and 1 T (circle).

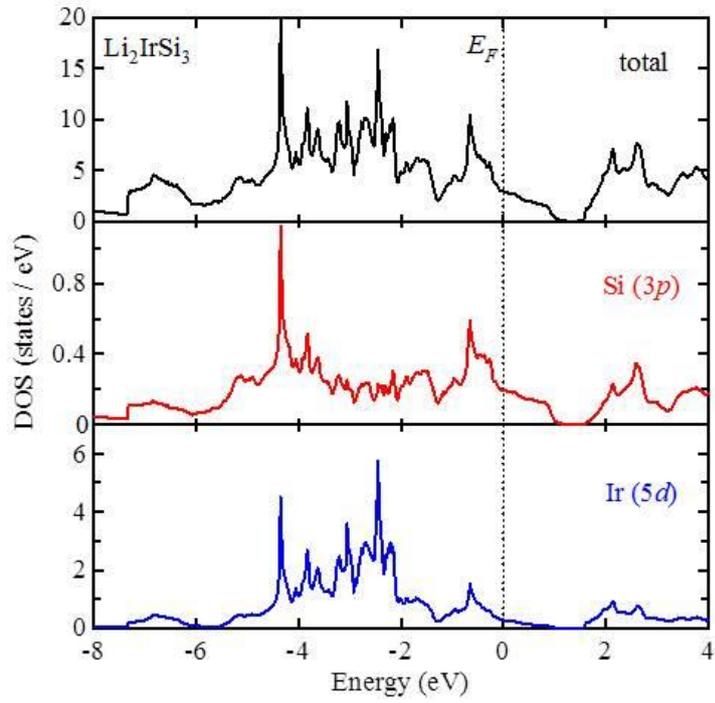

Fig. 3 Scalar-relativistic total density of states, symmetry resolved densities of Ir *d* and Si *p* states in $Li_2IrSi_3$, respectively.

Table I Structural parameters for $Li_2IrSi_3$ refined using Rietveld method from the synchrotron x-ray ($\lambda$ = 0.3998(2) Å) data at 293 K.

Crystal system; Hexagonal, Space group; $P6_3/mmc$ (No. 194), $a$ = 5.01762(1) Å, $c$ = 7.84022(1) Å, $Z$ = 2

| Atom | Site | $x$ | $y$ | $z$ | Occupancy | $B_{iso}$(Å$^2$) |
|---|---|---|---|---|---|---|
| Ir | 2$a$ | 0 | 0 | 0 | 1.0 | 0.288 |
| Si | 6$h$ | 0.3431(4) | 0.1716(2) | 3/4 | 1.0 | 0.378 |
| Li | 4$f$ | 1/3 | 2/3 | 0.5603(9) | 1.0 | 1.346 |

$R_{WP}$ = 13.42 %, $R_O$ = 9.82 %, $R_{Bragg}$ = 2.46 %, as defined in TOPAS 4.1